# Nanoscale spin-wave wake-up receiver

*Qi Wang, Thomas Brächer, Morteza Mohseni, Burkard Hillebrands, Vitaliy I. Vasyuchka, Andrii V. Chumak, Philipp Pirro*

*Fachbereich Physik and Landesforschungszentrum OPTIMAS, Technische Universität Kaiserslautern, Kaiserslautern, Germany*

We present the concept of a passive spin-wave device which is able to distinguish different radio-frequency pulse trains and validate its functionality using micromagnetic simulations. The information is coded in the phase of the individual pulses which are transformed into spin-wave packets. The device splits every incoming packet into two arms, one of which is coupled to a magnonic ring which introduces a well-defined time delay and phase shift. Since the time delay is matched to the pulse repetition rate, adjacent packets interfere in a combiner which makes it possible to distinguish simple pulse train patterns by the read-out of the time-integrated spin-wave intensity in the output. Due to its passive construction, this device may serve as an energy-efficient wake-up receiver used to activate the main receiver circuit in power critical IoT applications.

An ultra-low power wireless communication capability is needed to realize wireless sensor networks in the real-world scenarios. The most efficient way to reduce energy consumption is to leave the device off and wake it up asynchronously only when it is really needed. For this task, one needs a wake-up receiver which continuously monitors a channel, listens for a wake-up signal from other nodes and activates the device on stimulus [1]. Therefore, the power consumption as well as the footprint of the wake-up receiver are of crucial importance. Spin waves, a collective precessional motion of localized magnetic moments, can propagate at GHz frequencies in either conducting or insulating magnetic media with wavelength down to tens of nanometer and without any flow of electrons, reducing the energy consumption caused by Joule heating [2-5]. In general, spin waves, and their quanta magnons, are considered as potential candidate as carrier of information in future devices due to their low losses [6-9], short wavelengths [10-13] and high frequencies [14, 15]. Another advantage of spin waves is that due to their wave nature, they can be used for transfer and processing of information, which is coded in their amplitude and phase. Interference effects between different waves play a major role to harvest the full potential of the wave nature for data processing. A prominent example for the advantages of wave-based data processing is the majority gate, whose output state represents the majority of the input states [16, 17], and which is an especially compact and readily scalable wave-based device.

In this Letter, we design a magnonic wake-up receiver based on spin-wave interference. The basic concept of the magnonic wake-up receiver consists of Y-shaped splitter, a combiner and a coupled magnonic circular ring structure, as depicted in Fig. 1a. The device works in remanence. Radio frequency (RF) pulse trains from other wireless nodes will be converted into spin-wave pulse trains by a magnonic transducer in the input [18, 19]. The blue and red arrows show the flow path of the spin-wave packets in the magnonic wake-up receiver, which are split into two equal parts by a symmetric Y-shaped splitter. One of them is directly guided to the combiner region by the upper waveguide (blue channel). The other one is firstly transferred to the adjacent magnonic circular ring structure by dipolar coupling [5] (the red dashed area) and propagates along the ring and then couples back to the waveguide with an additional time delay $\Delta t$ and phase shift $\Delta \varphi$ (red channel). The time delay and phase shift can be controlled by adjusting the delay length $L_d$ [see Fig. 1b]. When the time delay $\Delta t$ is equal to the pulse period $\tau_p$ of the spin-wave pulse train, the back coupled spin wave in the red channel will completely interfere with the next packet of spin-wave pulse train in the blue channel at the combiner region. This way, a large output amplitude due to constructive interference is obtained only for a suitable correlation of the spin-wave phases in neighboring pulses of the input signal. Thus, a particular pulse sequence has to be used to create a strong output signal which serves to wake up the next connected device.

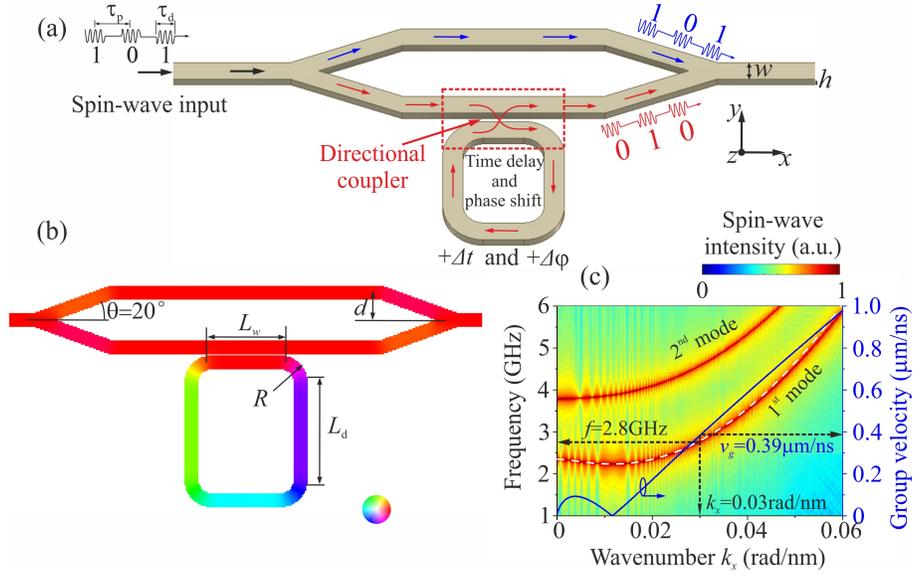

*Fig. 1 (a) Sketch of the spin-wave wake-up receiver consisting of a Y-shaped splitter, a combiner and a coupled circular ring structure. (b) Simulated ground state of the spin-wave wake-up receiver. The color code wheel represents the in-plane orientations of magnetization. (c) The simulated (color map) and theoretical (white dashed line) dispersion curve and theoretical (blue line) spin-wave group velocity.*

To investigate the spin-wave dynamics in the proposed structure, micromagnetic simulations were performed by Mumax3 [20]. The typical parameters of nanometer thick Yttrium Iron Garnet (YIG) were used in the simulations [9]: saturation magnetization $M_s = 1.4 \times 10^5$ A/m, exchange constant $A = 3.5$ pJ/m, and Gilbert damping $\alpha = 2 \times 10^{-4}$. The damping at the ends of the simulated structure was set to exponentially increase to 0.5 to prevent spin-wave reflection. In the real device, these waves could also be processed further in a larger magnonic network. The mesh was set to $10 \times 10 \times 50$ nm$^3$. The geometric sizes of the magnonic wake-up receiver are as follows: the widths of the waveguides are $w = 100$ nm, thicknesses are $h = 50$ nm, the minimum gap between the waveguide and the circular ring is 10 nm, the center to center distance of the waveguides is $d = 200$ nm. The angles of the Y-shaped splitter and combiner are $\theta = 20°$ and the outer radius of the corner is $R = 200$ nm as shown in Fig. 1b. The length where the waveguide and the circular ring structure are coupled to each other is adjusted to $L_w = 500$ nm as shown in Fig. 1b. By choosing a proper spin-wave frequency, all the incident wave energy is transferred from the waveguide to the magnonic ring structure [5] over this distance. Other frequencies can be used by changing $L_w$, the gap between the ring and the waveguide or the applied magnetic bias field.

Figure 1b shows the simulated ground state in which the color code shows the orientation of the magnetization. Please note that no external bias field is applied in our simulation. In the nanoscale waveguide, the static magnetization orients itself parallel to the $x$-direction spontaneously due to the strong shape anisotropy. In the circular ring structure, the static magnetization is the global vortex state where the magnetization orients itself along the ring. This global vortex state in the ring structure has already been reported experimentally [21,22] and it can be realized by applying a large external field along the $x$-direction to magnetize both, the waveguides and the ring structure in the $x$-direction and, consequently, slowly decreasing this field to zero. The global vortex then forms in the ring to minimize the total energy.

To determine the optimal parameters for the wake-up receiver, we first simulate the spin-wave dispersion in a single straight waveguide with the same parameters. A sinc function oscillation magnetic field $b_y = b_0 \text{sinc}(2\pi f_c t)$ (in $y$-direction) was applied in the center of waveguide over an area of 20 nm in length, with a maximum oscillation amplitude $b_0 = 1$ mT and a cut-off frequency of $f_c = 10$ GHz. Figure 1c shows the dispersion curves of the first two lowest spin-wave width modes of an isolated straight waveguide. The color map was obtained by micromagnetic simulations and the white dashed line for the lowest width mode was calculated using the analytic theory for fully unpinned boundary conditions [23]. The blue line shows the analytic group velocity of the lowest width mode. A frequency of $f = 2.8$ GHz ($k_x = 0.03$ rad/nm, $\lambda = 210$ nm, group velocity $v_g = 0.39$ μm/ns) has been chosen as working frequency. Then, only one wavenumber-frequency combination exists for the fundamental mode, avoiding the scattering between different width modes after the bent waveguides [24]. All the working parameters are indicated in Fig. 1(c).

As we mentioned above, the phase of the spin waves coupled back from the ring in the red channel in Fig. 1(a) can be adjusted by changing the delay length $L_d$ of the circular ring structure. Simulations with

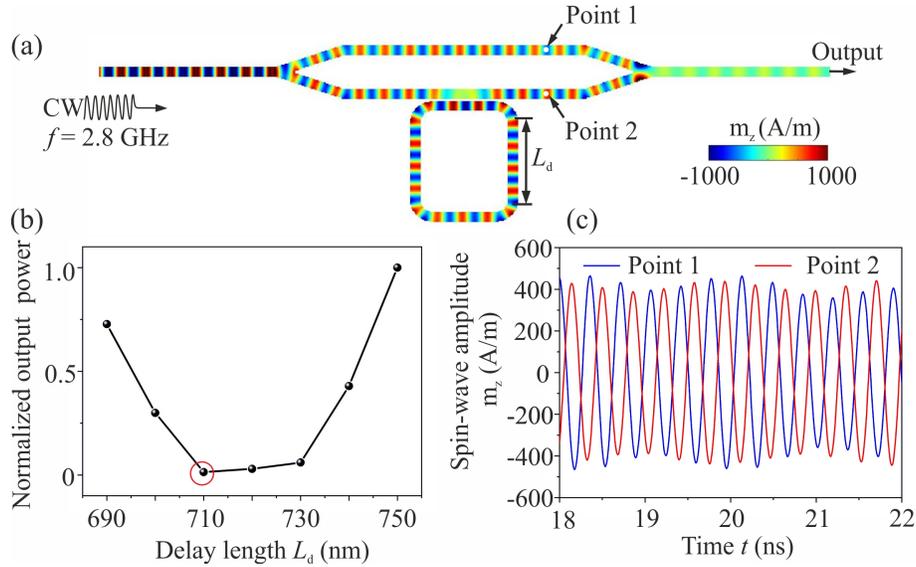

Fig. 2 (a) Snapshot of the out-of-plane component spin-wave amplitude $m_z$ for a continuous excitation source with frequency $f = 2.8$ GHz. (b) The normalized output power as a function of delay length $L_d$. (c) Spin-wave amplitudes at point 1 and 2 in the time range from 18 ns to 22 ns after excitation.

different delay lengths $L_d$ were performed with a continuous spin-wave excitation at frequency $f$. The normalized output power as a function of the delay length $L_d$ is shown in Fig. 2b. A minimum is found at $L_d = 710$ nm which corresponds to a destructive interference between the two branches. Figure 2(a) shows a snapshot of the out-of-plane component of the dynamics magnetization $m_z$, which is equivalent to the spin-wave amplitude, for a delay length of $L_d = 710$ nm. The obvious destructive interference which is observed in the combiner region implies that the circular ring structure introduces an effective phase shift of $\pi$ for the spin waves in the lower branch. Figure 2(c) shows the amplitudes of the spin-wave signal at point 1 and point 2 (marked in Fig. 2a) in the time range from 18 ns to 22 ns after excitation. From this, the phase shift of $\Delta\varphi = \pi$ is clearly visible.

To study the delay time caused by the coupled magnonic ring, a single spin-wave pulse of 5 ns duration ($\tau_d = 5$ ns) is applied. Figure 3(a) shows a snapshot of $m_z$ in which the spin-wave energy transferring between the waveguide and the magnonic ring is clearly observed. The output spin-wave amplitude as a function of time is shown in Fig. 3(b). Two clear peaks are found which correspond to the spin-wave packets exiting from the upper and the lower branch, respectively. The delay time of these two spin-wave packets is $\Delta t = 9.5$ ns (peak-to-peak distance). This value is slightly larger than the value of 8.6 ns estimated from the group velocity in Fig. 1. This is due to the fact that the group velocity in the corner of ring structure is smaller than the one in the straight waveguide [25]. Moreover, the amplitude of the second peak localized around 22 ns is smaller than that of the first one due to its longer propagation length caused by the coupled magnonic ring.

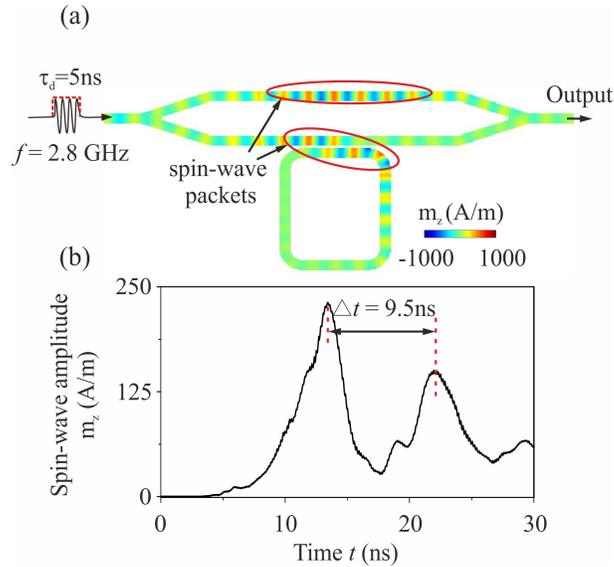

*Fig. 3 (a) Snapshot of the out-of-plane component spin-wave amplitude $m_z$. (b) The output spin-wave amplitude as a function of simulation time t.*

In this design, a RF pulse train that can be used to create a strong signal at the output is, for example, an alternating pulse train of logical 0 (phase 0) and logical 1 (phase π) as depicted in the top panel of Fig. 4(a). Since the induced time delay Δ*t* of the second pulse train (red wave in Fig. 4(a)) matches the period of the initial spin-wave pulse $\tau_p$ and shifts the phase of the wave packet by π, the logical bit number n+1 interferes always with the negated logical bit number n (n is integer). For the alternating pulse train 1010101010, this results in a constructive interference of the central spin-wave pulses, leading to a large envelop of the spin-wave intensity shown in the bottom panel of Fig. 4(a). Thus, with a RF to DC detector, the maximum possible voltage would be generated at the output. Other pulse sequences like the homogenous sequence 1111111111 create a much lower signal due to the destructive interference shown in Fig. 4(b). Please note that the output signal resulting from destructive interference is not zero because of the difference in amplitude of the two interfering spin waves. Thus, if a threshold voltage is defined, this scheme allows to trigger an action by a certain, specially shaped RF signal only. Please note also that the absolute value of the phase is not essential in this approach. Only the phase difference between neighboring bits is of interest, which is important since only this information can be transmitted between an arbitrarily placed sender and receiver. Another advantage is the passive property that does not require additional RF or DC sources as reference, which are hard to integrate on a chip. Please also note that a different geometric design can be used to encode other pulse sequences. For instance, the homogenous sequence 11111111 pulse train can also be used as a trigger signal by adjusting the delay length to meet the criterion in which the time delay Δ*t* is equal to the pulse period $\tau_p$ and the phase shift is zero.

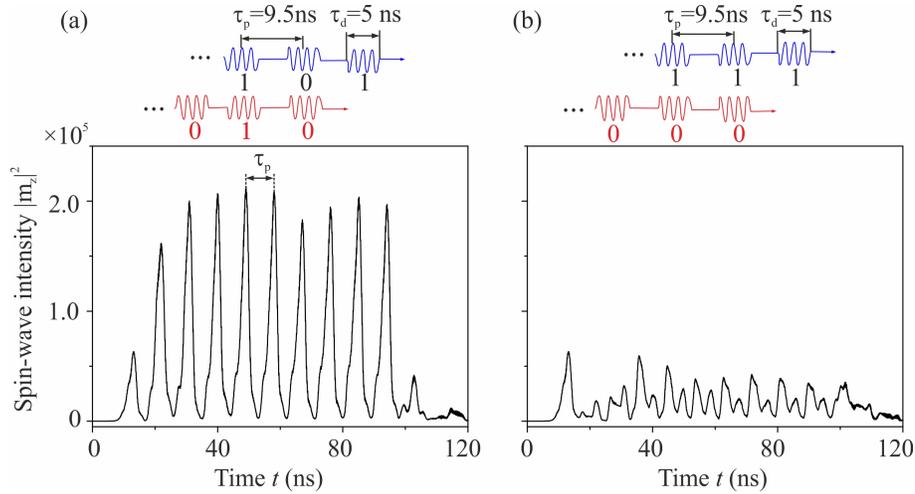

*Fig. 4 Top panel of (a) and (b): RF pattern recognition scheme (1010101010 and 1111111111): The red pulse train has been delayed by one pulse period $\tau_p$ and its phase is, consequently, shifted by Δφ = π. The bottom panel of (a) and (b): The simulated output spin-wave intensities for the two indicated input patterns.*

In summary, we have designed a planar nanoscale spin-wave wake-up receiver based on a simple wave interference effect and tested it by utilizing micromagnetic simulations. The wake-up receiver consists of a symmetric Y-shaped splitter, combiner and a coupled ring structure which acts as a phase shifter and delay line. The operational principle of the spin-wave wake-up receiver is demonstrated. Relying on spin-wave interference, a dedicated RF pulse train can be used to create a strong signal at the output to wake up the next connected device. The presented device operates intrinsically passive and does not require any additional RF sources as a reference.

**Acknowledgement:** Financial support by DFG within project B01 of the Transregional Collaborative Research Centre (SFB/TRR) 173 "Spin+X" and ERC Starting Grant 678309 MagnonCircuits and the EU Horizon 2020 research and innovation programme within the FET-OPEN project CHIRON (contract number 801055) is gratefully acknowledged.

**Reference**

1. G. U. Gamm, M. Sippel, M. Kostic, and L. M. Reindl, Sensor Networks and Information Processing (ISSNIP), Brisbane, QLD, Australia, 7–10 December 2010; pp. 121–126.
2. A. Khitun, M. Bao, and K. L. Wang, *J. Phys. D Appl. Phys.* **43**, 264005 (2010).
3. B. Lenk, H. Ulrichs, F. Garbs, and M. Münzenberg, *Phys. Rep.* **507**, 107 (2011).
4. A. V. Chumak, V. I. Vasyuchka, A. A. Serga, and B. Hillebrands, *Nat. Phys.* **11**, 453 (2015).
5. Q. Wang, P. Pirro, R. Verba, A. Slavin, B. Hillebrands, and A. V. Chumak, *Sci. Adv.* **4**, e1701517 (2018).
6. Y. Kajiwara, K. Harii, S. Takahashi, J. Ohe, K. Uchida, M. Mizuguchi, H. Umezawa, H. Kawai, K. Ando, K. Takanashi, S. Maekawa, and E. Saitoh, *Nature* **464**, 262 (2010).
7. P. Pirro, T. Brächer, A. V. Chumak, B. Lägel, C. Dubs, O. Surzhenko, P. Görnert, B. Leven, and B. Hillebrands, *Appl. Phys. Lett.* **104**, 012402 (2014).
8. H. Yu, O. d'Allivy Kelly, V. Cros, R. Bernard, P. Bortolotti, A. Anane, F. Brandl, R. Huber, I. Stasinopoulos, and D. Grundler, *Sci. Rep.* **4**, 6848 (2014).
9. C. Dubs, O. Surzhenko, R. Linke, A. Danilewsky, U. Brückner, and J. Dellith, *J. Phys. D Appl. Phys.* **50**, 204005 (2017).
10. S. Wintz, V. Tiberkevich, M. Weigand, J. Raabe, J. Lindner, A. Erbe, A. Slavin, and J. Fassbender, *Nat. Nano.*, **11**, 948 (2016).
11. H. Yu, O. d'Allivy Kelly, V. Cros, R. Bernard, P. Bortolotti, A. Anane, F. Brandl, F. Heimbach, and D. Grundler, *Nat. Commun.* **7**, 11255 (2016).
12. T. Brächer, M. Fabre, T. Meyer, T. Fischer, S. Auffret, O. Boulle, U. Ebels, P. Pirro, and G. Gaudin, *Nano Lett.* **17**, 7234 (2017).

13. S. J. Hämäläinen, F. Brandl, K. J. A. Franke, D. Grundler, and S. Van Dijken, *Phys. Rev. Appl.* **8**, 014020 (2017).
14. T. Kampfrath, A. Sell, G. Klatt, A. Pashkin, S. Mährlein, T. Dekorsy, M. Wolf, M. Fiebig, A. Leitenstorfer, and R. Huber, *Nature Photon.* **5**, 31 (2011).
15. T. Balashov, P. Buczek, L. Sandratskii, A. Ernst, and W. Wulfhekel, *J. Phys.: Condens. Matter*, **26**, 394007 (2014).
16. S. Klingler, P. Pirro, T. Brächer, B. Leven, B. Hillebrands, and A. V. Chumak, *Appl. Phys. Lett.* **105**, 152410 (2014).
17. T. Fischer, M. Kewenig, D. A. Bozhko, A. A. Serga, I. I. Syvorotka, F. Ciubotaru, C. Adelmann, B. Hillebrands, and A. V. Chumak, *Appl. Phys. Lett.* **110**, 152401 (2017).
18. T. Kimura, Y. Otani, T. Sato, S. Takahashi, and S. Maekawa, *Phys. Rev. Lett.*, **98**, 156601 (2007).
19. S. Cherepov, P. K. Amiri, J. G. Alzate, K. Wong, M. Lewis, P. Upadhyaya, J. Nath, M. Bao, A. Bur, T. Wu, G. P. Garman, A. Khitun, and K. L. Wang, *Appl. Phys. Lett.* **104**, 082403 (2014).
20. A. Vansteenkiste, J. Leliaert, M. Dvornik, M. Helsen, F. Garcia-Sanchez, and B. Van Waeyenberge, *AIP Adv.* **4**, 107133 (2014).
21. M. Steiner and J. Nitta, *Appl. Phys. Lett.* **84**, 939 (2004).
22. Y. Ren and A. O. Adeyeye, *J. Appl. Phys.* **105**, 063901 (2009).
23. Q. Wang, B. Heinz, R. Verba, M. Kewenig, P. Pirro, M. Schneider, T. Meyer, B. Lägel, C. Dubs, T. Brächer, and A. V. Chumak, arXiv:1807.01358.
24. P. Clausen, K. Vogt, H. Schultheiss, S. Schäfer, B. Obry, G. Wolf, P. Pirro, B. Leven, and B. Hillebrands, *Appl. Phy. Lett.* **99**, 162505 (2011).
25. N. Kumar, G. Venkat and A. Prabhakar, *IEEE Trans. Magn.* **50**, 1300306 (2014).